\documentstyle[12pt,amstex,amssymb,righttag]{article}
\begin{document}
\renewcommand{\thefootnote}{\fnsymbol{footnote}}

\thispagestyle{empty}

\vspace*{-1cm}
\begin{center}
{\Large \bf Some Comments of Vlasov-Liouville Equation \\ and its Relation with
Gravitational Field}

\vspace{3mm}
by\\
\vspace{3mm}
{\sl Carlos Pinheiro$^{\ddag}$\footnote{fcpnunes@@cce.ufes.br/maria@@gbl.com.br}
} 
and 
{\sl F.C. Khanna$^{+}$\footnote{khanna@@phys.ualberta.ca}}

\vspace{3mm}
$^{\ddag}$Universidade Federal do Esp\'{\i}rito Santo, UFES.\\
Centro de Ci\^encias Exatas\\
Av. Fernando Ferrari s/n$^{\underline{0}}$\\
Campus da Goiabeiras 29060-900 Vit\'oria ES -- Brazil.\\

$^{+}$Theoretical Physics Institute, Dept. of Physics\\
University of Alberta,\\
Edmonton, AB T6G2J1, Canada\\
and\\
TRIUMF, 4004, Wesbrook Mall,\\
V6T2A3, Vancouver, BC, Canada.
\end{center}

\vspace{3mm}
\begin{center}
Abstract
\end{center}

We discuss here the possibility to write the Liouville-Vlasov
equation for the Wigner-function of a spinor field coupled to a
gauge field with field strength tensor $F^{\mu\nu}$ in a curved
space-time versus a local Lorentz manifold (introduction of local
Lorentz coordinates) with an appropriate definition of a covariant
derivative carried out using a spin connection $B_{\mu}^{ab}(x)$.

\newpage
\section{Introduction}\setcounter{footnote}{0}
\paragraph*{}

The general relativity is invariant under
general coordinate transformations like $x\rightarrow x' = x+\xi (x)$.
One can still define a convenient covariant derivative under a
general transformation of coordinate as ${\cal
D}_{\mu}\xi_{\nu}=-\partial_{\mu}\xi_{\nu}+\Gamma_{\mu\nu}^{\lambda}\xi_{\lambda}$
where $\xi_{\mu}(x)$ is the infinitesimal parameter and
$\Gamma_{\mu\nu}^{\lambda}$ is the Christoffel symbol or ``connection''.

The infinitesimal variation of metric under the coordinate
transformation can be written as $\delta g_{\mu\nu}(x)=-{\cal
D}_{\mu}\xi_{\mu}(x) - {\cal D}_{\nu}\xi_{\mu}(x)$.

It's easy to verify that for weak field approximation,
$g_{\mu\nu}(x)=\eta_{\mu\nu}+kh_{\mu\nu}$ with a redefinition of the 
$\xi_{\mu}(x)$ parameter. We obtain the gauge transformation
imposed on the field $h_{\mu\nu}(x)$ such as infinitesimal
version in the Minkowsky space under GCT's (general transformation
of coordinates).

How do we consider the interaction between gravitation and other
fields? The general recipe is to introduce new fields and to define
an appropriate covariant derivative that includes such an interaction.

In general there are two ways to follow: Firstly, we can use the
coupling between gravitation and other fields. In this case the
gravitational field is only a background for other fields. The
scenario is geometrical as for gravity which acts as a background
field. There is no ``interaction'' as is the
case for the electromagnetic theory, quantum  chromodynamics or even in the
perturbative quantum gravity. On the other hand we can introduce a
flat space time and attack the problem using the perturbative
approach. If we wish to interpret the fields as particles
associated with that field we need to introduce a local Lorentz
manifold to bring in a specific interactions. It is true, in particular if
we wish to analyse the interaction between gravitation and spinor fields.
The interaction cannot be introduced directly in curved space time for
the spinor field. It's necessary to look for a local Lorentz
manifold rather than the global manifold as in space time. 

This point is emphasized, in particular, for the special case of
Liouville-Vlasov equation. We believe that the correct way to
introduce the effect of gravitational field in that
equation is by defining an appropriate covariant derivative in a
local Lorentz manifold against the point of view that one can define
the gravitational effect on Vlasov equation directly in curved space
time. The latter is written as follows. 

We analyse the interaction between gravitation and spinor fields
using the vierbein field and the local Lorentz group as our support.
As a second example we present the coupling between matter-Maxwell
and gravitation fields. The following example is the $U(1)$ 
gauge group generalized to the case of, $SO(N)$ model
coupled to gravity and spinor-vector field (Rarita-Schwinger). In all
cases the interaction is seen on a local Lorentz manifold where it is
possible to find a spin connection, and in some cases, to find the
``torsion'' associated with such a model. 

Finally, the Liouville-Vlasov equation is discussed and the effect of
gravitational field on that equation is analysed. We argue here that
it is impossible to define a covariant derivative for the Vlasov
equation if $\psi (q,p)$  means the Wigner function associated with
the spinor field. The fundamental reason for it is that the symmetry
group of general relativity is incompatible with the presence of
spinors [2]; but we can do that only in a local Lorentz manifold
where both the Vierbein and spin connection are defined. We start
by remembering that the concept of spin  $1/2$ field makes sense only
in a tangent flat space [4,5]. One can write the action for
interaction between Dirac's field and gravitation as
\begin{equation}
S=\int d^4x \ e\left[e^{\mu a}\bar{\psi} i\gamma_a \ D_{\mu}\psi - m \bar{\psi}\psi\right]
\end{equation}
where
\begin{equation}
\gamma^{\mu} = e^{\mu}_a (x) \gamma^a
\end{equation}
and
\begin{equation}
D_{\mu}\psi_{\alpha} = \partial_{\mu}\psi_{\alpha} + \frac{i}{8} B_{\mu}^{ab}
[\gamma_a\gamma_b]_{\alpha\beta} \psi_{\beta}
\end{equation}
with $e^{\mu}_a$ are the Vierbein field with two indices; one space time
index a, $\mu$, varying from zero to three and a local Lorentz group
index, a, of a local manifold. The $\gamma^{\mu}$ are the Dirac's matrices,
$\gamma^a$ meaning the local Dirac matrices and $B^{ab}_{\mu}$ being
the spin connection.

The interaction between gravitation and fermionic field can be seen
immediately from eq. (3) because of the appropriate definition
of a covariant derivative $D_{\mu}$ for a local Lorentz group.

The metric can be written as 
\begin{equation}
g_{\mu\nu} = e_{\mu}^ae^b_{\nu} \eta_{ab}\ .
\end{equation}
Thus, the ``{\it e}'' in eq. (1) means the determinant of the
metric. {\it The Dirac equation coupled to gravitation}   is given by
\begin{equation}
(e^{\mu a}i\gamma_a D_{\mu}-m) \psi (x) = 0
\end{equation}
The spin connection in (3) can be completly fixed by the coefficient of
non holomicity as
\begin{equation}
B^{ab}_{\mu} = -\ \frac{1}{2} e^c_{\mu} (\Omega_{cda}+\Omega_{acd}-\Omega_{dac})
\end{equation}
where 
\begin{equation}
\Omega_{cda} = e^{\rho}_c e^{\sigma}_j (\partial_{\rho}e_{\sigma a}-
\partial_{\sigma}e_{\rho a})
\end{equation}
The equation for $e^a_{\mu}$ field, $R_{\mu a}-\frac{1}{2}e_{\mu a}R=0$ fixes
$e_{\mu}^a (x)$ exactly; then we have that $B_{\mu}^{ab}$ shall be fixed too.

The gravitational degree of freedom is carried by Vierbein. All
information about spins from gravitational field will be given by $e_{\mu}^a(x)$.

One may see clearly that the kinetic term of the fermionic lagrangean
contributes to the equation of motion for the spin connection and we can
show that the presence of the fermionic field generates torsion [4,5]
given as
\begin{equation} 
T_{\mu\nu}^a \sim D_{\mu} (e e^{\mu}_{[a}e^{\nu}_{b]}) \sim
\bar{\psi}\gamma [\gamma ,\gamma ]\psi \ .
\end{equation}
where $[a,b]$ here means symmetrization and the right hand side represents
the fermionic density.

\section*{The interaction between matter-Maxwell and gravitational fields}
\paragraph*{}

The action for interaction between matter-Maxwell and gravitational
fields is given as
\begin{equation}
S^{Maxwell}_{Mat-Grav} = \int d^4x \ e[e^{\mu a} \bar{\psi} i\gamma_a
\nabla_{\mu} \psi - m \bar{\psi}\psi ]
\end{equation}
where $\nabla_{\mu}$ is a covariant derivative of gravitation and
the gauge is given by
\begin{equation}
\nabla_{\mu}\psi = D_{\mu} + ig q A_{\mu}
\end{equation}
with $D_{\mu}$ being the covariant derivative as in eq. (3), $A_{\mu}(x)$
is the vector potential, $q$ is a coupling constant associated with
the $U(1)$ symmetry and $g$ is another coupling constant linking the
local Lorentz group. The action is invariant under local Lorentz
group, GCT's (transformation coordinates group) and $U(1)$ symmetry simultaneously.

A term is needed that gives the dynamics for the gauge degree of  freedom
(Maxwell term); so the complete action which couples
Maxwell-Dirac-gravitation is written as
\begin{equation}
S^{Maxwell}_{Dirac-Grav} = \int d^4x \ e[-\frac{1}{4}
F_{\mu\nu}F^{\mu\nu} + e^{\mu a}\bar{\psi} i\gamma_a \nabla_{\mu}
\psi - m\bar{\psi}\psi ]
\end{equation}  

\section*{The $U(1)$ case generalized}
\paragraph*{}

Consider the scalar and spinor field matter fields  coupled to
gravitation and Maxwell field. The lagrangean is given as
\begin{equation}
{\cal L} =
(\partial_{\mu}\varphi^{\ast}\partial^{\mu}\varphi -m^2\varphi^{\ast}\varphi
) + \bar{\psi} (i\gamma^{\mu}\partial_{\mu}-M) \psi +
\frac{\lambda}{4} (\varphi^{\ast}\varphi )^2\ .
\end{equation}

To obtain a covariant lagrangean the convenient covariant derivative
is introduced as $\partial_{\mu}\rightarrow \nabla_{\mu}$, where 
\begin{equation}
\nabla_{\mu}\psi =\left(\partial_{\mu} + \frac{i}{8}
B^{ab}_{\mu}[\gamma_a,\gamma_b] + ig QA_{\mu}(x)\right)\psi
\end{equation}
Thus, the complete action for Maxwell, matter and
gravitational field is 
\begin{equation} 
S^{Maxwell}_{mat-grav} = \int d^4x \
e\left[(\nabla_{\mu}\varphi^{\ast})(\nabla^{\mu}\varphi ) -
V(\varphi^{\ast},\varphi )\right] + e \ e^{\mu a}\bar{\psi}
(i\gamma_a\nabla_{\mu}\psi - M) \psi
\end{equation}   
where $V(\varphi^{\ast}\varphi ) = m^2\varphi^{\ast}\varphi$ is the
potential term and $Q$ means the generalized charge.

The equation of motion is immediately obtained as
\begin{equation}
\nabla^{\mu}\nabla_{\mu} \varphi - m^2 \varphi + \cdots = 0
\end{equation}
and the dynamics is verified on local Lorentz manifold again.

\section*{Model $SO(N)$ coupled to gravity}
\paragraph*{}

The interest in this model is formal. There appear magnetic monopoles
of t'Hooft-Polyakov coupling to gravitational field. 

Suppose that $\varphi_a$ are scalars fields in the representation of
$SO(N)$ and $a=1,2,3\cdots N$.

Thus,
\begin{equation}
\varphi_a{}' = R_{ab}\varphi_b
\end{equation}
where $R_{ab}$ is the  transformation matrix and it satisfies 
the condition $R^+R=1$.

The lagrangean satisfying the invariance under global $SO(N)$ is
given by
\begin{equation}
{\cal L} = \frac{1}{2} \partial_{\mu}
\varphi_a\partial^{\mu}\varphi_a - \frac{1}{2} m^2 \varphi_a\varphi_a
\ .
\end{equation}
the $SO(N)$ will be calibrated by equation (16) and we need to change
again the derivative $\partial_{\mu}$ to a new covariant derivative
$\nabla_{\mu}$ written as
\begin{equation}
\nabla_{\mu}\varphi_a = (\partial_{\mu}\delta_{ab}+ig A_{\mu}^I
(G_I)_{ab}) \varphi_b
\end{equation}
where $G_I$ are generators of $SO(N)$ group and $A_{\mu}^I(x)$ are
the non abellian vector potentials. The number of generators being
written as $I=1\cdots \displaystyle{\frac{N(N-1)}{2}}$.

The complete action for this case can be written as
\begin{eqnarray}
&& S^{Y.M}_{mat-grav} = \int d^4x \ e\left[-\frac{1}{4} F_{\mu\nu I}
F^{\mu\nu I} + \frac{1}{2} \nabla_{\mu}\varphi_a\nabla^{\mu}\varphi_a
- \frac{1}{2} m^2 \varphi_a\varphi_a + \right.\nonumber \\
&&\left. -\frac{\lambda}{4!} (\varphi_a\varphi_b)^2 + fR \varphi_a\varphi_a\right]
\end{eqnarray} 
where $f$ means the dimensionless coupling constant, $R$ in the last
term is the scalar curvature and $F_{\mu\nu I}$ represents the field
strength given by 
\begin{equation}
F_{\mu\nu I} = \partial_{\mu}A_{\nu I} - \partial_{\nu}A_{\mu I} + g
f_{IJK} A_{\mu J}A_{\nu K}
\end{equation}

\section*{Spinor-vector fields}
\paragraph*{}

We define now $\psi_{a\alpha}$ as a spinor-vector field with $\alpha
= 1,2,3,4$ and $a=0,1,2,3$. The transformation law for
Rarita-Schwinger field can be written as
\begin{equation}
\delta\psi_{a\alpha} = w_a^b\psi_{b\alpha} + \frac{1}{2} w^{mn}
[\gamma_m,\gamma_n]_{\alpha\beta} \psi_{\alpha\beta}
\end{equation}   
where $w^b_a$ are continuously varying parameters. The first part on
right hand side being responsible for spin-1 and the second part being
associated with spin field $1/2$ and $3/2$. 

The spin $1/2$ can be eliminated consistently by a symmetry of the
free lagrangean [5]. Thus, $\psi_{a\alpha}$ shall describe a pure
spin $3/2$.

It appears in supergravity [5] as a fermionic mediator of the
gravitational interaction. 

The lagrangean which describes a free spinor-vector field can be
given as
\begin{equation}
{\cal L}_{R.S.} = \frac{1}{2} \varepsilon^{\mu\nu\rho\sigma}
\bar{\psi}_{\mu} \gamma_5\gamma_{\nu} \partial_{\rho} \psi_{\sigma}
\end{equation}
with $\varepsilon^{\mu\nu\rho\sigma}$ being the Levi-Civita tensor
and $\gamma_5=\gamma_0\gamma_1\gamma_2\gamma_3$ is a Dirac's matrix.
We can write eq. (22) in the component form as
\begin{equation}
{\cal L}_{R.S.} = \frac{1}{2} \varepsilon^{\mu\nu\rho\sigma}
\bar{\psi}_{\mu\alpha} (\gamma_5\gamma_{\nu})_{\alpha\beta} \partial_{\rho}\psi_{\sigma\beta}
\end{equation}
The local invariance $U(1)$ that permits the elimination of spin
$1/2$ components is given as    
\begin{equation}
\delta\psi_{\mu\alpha} = \partial_{\mu}\chi 
\end{equation}
where $\chi$ is a spinor field. As in the case of the fermionic field the
Rarita-Schwinger field generates torsion [5].

The interaction between gravitation and the spinor-vector field is
shown following the same recipe as in earlier cases so that 
\begin{equation}
{\cal L}_{R.S.}^{gravity} = \frac{e}{2} \varepsilon^{\mu\nu\rho\sigma}
\bar{\psi}_{\mu\alpha} (\gamma^5\gamma_{\nu})_{\alpha\beta} D_{\rho}
\psi_{\sigma \beta}
\end{equation}
where
\begin{equation}
D_{\rho}\psi_{\sigma\beta} = \partial_{\rho}\psi_{\sigma\beta} +
\frac{i}{8} B_{\rho}^{ab} [\gamma_a,\gamma_b]_{\beta\gamma} \psi_{\sigma\gamma}
\end{equation}
It is clear that the only difference between eq. (25) and the other cases
is that here $\psi$ has two indices. One can show again that the
spinor-vector field generates torsion again as $T_{\mu\nu}^a\sim
\bar{\psi}\gamma_5\gamma [\gamma ,\gamma ]\psi$ forming a spinor
condensity exactly the same way as given in (8).

\section*{The Liouville-Vlasov equation with the gravitational interaction}
\paragraph*{}

Finally, we obtain the Liouville-Vlasov equation [3] for the Wigner
function of a ``spinor field'' coupled to a gauge field with the field
strength tensor $F^{\mu\nu}$ as 
\begin{equation}
(-p_{\mu}F^{\mu\nu} D_{\nu} + P^{\mu}D_{\mu}) \psi (q,p) = 0
\end{equation}
where $\psi (q,p)$ is the Wigner function. Here $D_{\mu}$ cannot be a
covariant derivative including the curved space-time spin connection
unlike the case discussed in [1,3].

The reason for this is that the  coordinate tranformation 
group of general relativity is incompatible with the presence of
spinor field and so, incompatible with the appearance of spin
connection $B_{\mu}^{ab}$ in curved space-time [2,4].

We can introduce an appropriate covariant derivative in eq. (27) the same
way as in preceding cases. So, considering the tangent space or local
Lorentz group we do the same use of eq. (3) and in all analyzed cases,
obtained for $D_{\mu}$ in eq. (27) a covariant derivative like 
\begin{equation}
D_{\mu}\psi = \partial_{\mu}\psi + \frac{i}{8} B^{ab}_{\mu}
[\gamma_a,\gamma_b] \psi
\end{equation}
where $\psi$ is the Wigner function associated with a spinor
field coupled to a gauge field in a local Lorentz manifold (tangent
space). It is well-known that [3] $\psi^{\ast}(q,p)\psi (q,p)$ can be
interpreted as the classical distribution function in the
relativistic phase space.

We observe that if we wish to include the real effects from
gravitational field in an approach of general relativity together
with the 
kinetic theory some troubles will appear. For example:
it is clear  [3] ``that the treatment of transport theory in a curved
space-time background is hidered by the fact that a definition of
the Wigner function $\psi (q,p)$ depends on the use of the Fourier
transform of a space-time correlation function with a translated argument''.

Fourier transformation (nor translation) are globally available in
general curved spacetime. On the other hand the symmetry group  of
general relativity is not compatible with the presence of spinors or
spin connection field [2] if the Wigner function
of a spinor field in a curved space time has to be found.

However, as discussed in [3], the construction of Wigner function $\psi (q,p)$
is still meaningful if carried out in the tangent space as in local
Lorentz manifold as applied in our case. The problem is that
the correlation between two points on that manifold is by an exponential
map [3]. Only this way, we can introduce the fermionic field
interacting with the gravitation and to define an appropriate
covariant derivative containing the spin connection.

\subsection*{Acknowledgements:}

\paragraph*{}
I would like to thank the Department of Physics, University of
Alberta for their hospitality. This work was supported by CNPq
(Governamental Brazilian Agencie for Research.

I would like to thank also Dr. Don N. Page for his kindness and attention
with  me at Univertsity of Alberta and Dr. Ademir Santana from
Federal University of Bahia by fruitful discussions.

\end{document}